\documentstyle[epsf,psfig,12pt]{article}
\renewcommand{\baselinestretch}{1.05}

\begin{document}
\def\be{\begin{eqnarray}}
\def\en{\end{eqnarray}}
\def\up{\uparrow}
\def\dw{\downarrow}
\def\non{\nonumber}
\def\la{\langle}
\def\ra{\rangle}
\def\ep{\varepsilon}
\def\vma{{_{V-A}}}
\def\vpa{{_{V+A}}}
\def\m{\hat{m}}
\def\fp{{f_{\eta'}^{(\bar cc)}}}
\def\half{{{1\over 2}}}
\def\pr{{\sl Phys. Rev.}~}
\def\prl{{\sl Phys. Rev. Lett.}~}
\def\pl{{\sl Phys. Lett.}~}
\def\np{{\sl Nucl. Phys.}~}
\def\zp{{\sl Z. Phys.}~}

\font\el=cmbx10 scaled \magstep2
{\obeylines
\hfill IP-ASTP-03-97
\hfill NTU-TH-97-08
\hfill September, 1997}

\vskip 1.5 cm

\centerline{\large\bf Charmless $B$ Decays to $\eta'$ and $\eta$}
\medskip
\bigskip
\medskip
\centerline{\bf Hai-Yang Cheng}
\medskip
\centerline{ Institute of Physics, Academia Sinica}
\centerline{Taipei, Taiwan 115, Republic of China}
\medskip
\centerline{\bf B. Tseng}
\medskip
\centerline{Department of Physics, National Taiwan University}
\centerline{Taipei, Taiwan 106, Republic of China}
\bigskip
\bigskip
\bigskip
\centerline{\bf Abstract}
\bigskip
{\small   
Exclusive charmless $B$ decays to $\eta'$ and $\eta$ are studied 
in the standard effective Hamiltonian approach. The effective coefficients 
that take into account nonfactorizable effects in external and internal 
$W$-emission amplitudes are fixed by experiment.
Factorization is applied to the matrix elements of QCD,
electroweak penguin operators and next-to-leading corrections to penguin
Wilson coefficients, which are renormalization-scheme dependent, are included, 
but the resultant amplitude is 
renormalization-scheme and -scale independent. We show that the unexpectedly
large branching ratio of $B^\pm\to \eta' K^\pm$ measured by CLEO can be 
explained by the Cabibbo-allowed internal $W$-emission process $b\to c\bar cs$ 
followed by a conversion of the $c\bar c$ pair into the $\eta'$ 
via gluon exchanges characterized by a decay
constant of order $-50$ MeV. Implications are discussed. The mechanism
of the transition $c\bar c\to\eta'$, if correct, will enhance the
decay rates of $B\to \eta' K^{*}$ by more than an
order of magnitude. However, it plays only a minor role to the rare decays
$B\to \eta'(\eta)\pi(\rho)$. The predicted branching ratio of $B^\pm\to
\eta\pi^\pm$ for positive $\rho$ is marginally ruled out by experiment, 
indicating that the favored Wolfenstein parameter $\rho$ is negative.

}

\pagebreak
 
{\bf 1.}~~The inclusive and exclusive rare $B$ decays to $\eta'$ have recently 
received a great deal of attention. The CLEO collaboration 
has reported very large branching ratios for inclusive $\eta'$
production \cite{CLEO1}
\be
{\cal B}(B^\pm\to\eta' X_s;~2.0~{\rm GeV}\leq p_{\eta'}\leq 2.7~{\rm GeV})=
(6.2\pm 1.6\pm 1.3)\times 10^{-4},
\en
and for the exclusive decay $B\to\eta'K$ \cite{CLEO2}
\be
{\cal B}(B^\pm\to\eta'K^\pm) &=& \left(7.1^{+2.5}_{-2.1}\pm 0.9\right)
\times 10^{-5}, \non\\
{\cal B}(\stackrel{_{_{(-)}}}{B}\!\!{^0}\to\eta' \stackrel{_{_{(-)}}}{K}\!\!
{^0}) &=& \left(5.3^{+2.8}_{-2.2}\pm 1.2\right)\times 
10^{-5}.
\en
These results are abnormaly large; for example, the branching ratio of
$B^\pm\to \eta'K^\pm$ is theoretically estimated to be of order $(1-2)\times 
10^{-5}$ in the standard approach, which is too small by $2.5\sigma$ compared
to experiment. It is natural to speculate that the unexpected
observation has something to do with the unique and special feature of
the $\eta'$ meson.

    Several mechanisms have been advocated to explain the copious $\eta'$
production in the inclusive process $B\to\eta'+X$: (i) the penguin mechanism
$b\to s+g^*$ followed by $g^*\to \eta'+g$ via the QCD anomaly \cite{Soni,Hou}.
Whether or not the standard $b\to s+g^*$ penguin alone is adequate to
explain the data (1) is still under debate, (ii) the Cabibbo-allowed
process $b\to (c\bar c)_1+s$ (the subscript 1 denotes color singlet),
followed by the transition $(c\bar c)_1\to {\rm two~gluons}\to\eta'$. It
was argued in \cite{Halp1} that the decay constant $f_{\eta'}^{(c\bar c)}$,
to be defined below, should be as large as 140 MeV in order to explain
the inclusive $\eta'$ data solely by this mechanism, and (iii) the process 
$b\to(c\bar c)_8+s$, analogous to (ii) but with the $c\bar c$ pair in a 
color-octet state, followed by $(c\bar c)_8\to\eta'+X$ \cite{Chao}.

  In this Letter we will study the exclusive charmless $B$ decays to 
$\eta'$ and $\eta$ and show that the CLEO measurement of $B^\pm\to\eta' K^\pm$
is indeed larger than what expected from the standard approach based on 
the effective Hamiltonian and factorization. We then proceed to suggest that
two of the aforementioned mechanisms, though important for inclusive $\eta'$
production, do not play any significant role to the two-body decay $B\to \eta' 
K$ and that a reasonable amount of the charm content of the $\eta'$ arising 
from the transition $c\bar c\to\eta'$, corresponding to $|\fp|\sim 50$ MeV,
suffices to account for the data. Many interesting implications due to 
this mechanism are discussed.

\medskip
{\bf 2.}~~The hadronic rare $B$ decays and their CP violation have been 
studied in 
detail in \cite{Chau1,Chau2} (see also \cite{Dean,Kramer,Du}). The relevant 
effective $\Delta B=1$ weak Hamiltonian is
\be
{\cal H}_{\rm eff}(\Delta B=1) = {G_F\over\sqrt{2}}\Big[ V_{ub}V_{uq}^*(c_1
O_1^u+c_2O_2^u)+V_{cb}V_{cq}^*(c_1O_1^c+c_2O_2^c)
-V_{tb}V_{tq}^*\sum^{10}_{i=3}c_iO_i\Big]+{\rm h.c.},
\en
where $q=u,d,s$, and
\be
&& O_1^u= (\bar ub)_\vma(\bar qu)_\vma, \qquad\qquad\quad~~O_1^c = (\bar cb)_
\vma(\bar qc)_\vma,   \non \\
&& O_2^u = (\bar qb)_\vma(\bar uu)_\vma, \qquad \qquad \quad~~O_2^c = 
(\bar qb)_\vma(\bar cc)_\vma,   \non \\
&& O_3=(\bar qb)_\vma\sum_{q'}(\bar q'q')_\vma, \qquad \qquad O_4=(\bar q_
\alpha b_\beta)_\vma\sum_{q'}(\bar q'_\beta q'_\alpha)_\vma,   \non \\
&& O_5=(\bar qb)_\vma\sum_{q'}(\bar q'q')_\vpa, \qquad\qquad O_6=(\bar q_
\alpha b_\beta)_\vma\sum_{q'}(\bar q'_\beta q'_\alpha)_\vpa,   \non \\
&& O_7={3\over 2}(\bar qb)_\vma\sum_{q'}e_{q'}(\bar q'q')_\vpa,  \qquad
O_8={3\over 2}(\bar q_\alpha b_\beta)_\vma\sum_{q'}e_{q'}(\bar q'_\beta
q'_\alpha)_\vpa,   \non \\
&& O_9={3\over 2}(\bar qb)_\vma\sum_{q'}e_{q'}(\bar q'q')_\vma,  \qquad
O_{10}={3\over 2}(\bar q_\alpha b_\beta)_\vma\sum_{q'}e_{q'}(\bar q'_\beta
q'_\alpha)_\vma,
\en
with $(\bar q_1q_2)_{_{V\pm A}}\equiv\bar q_1\gamma_\mu(1\pm \gamma_5)q_2$. 
In Eqs.~(3) and (4), $O_3-O_6$ are QCD penguin operators and $O_7-O_{10}$ 
are electroweak penguin operators.

The Wilson coefficients $c_i(\mu)$ in Eq.~(1) have been evaluated at the
renormalization scale $\mu\sim m_b$ to the next-to-leading order. Beyond the
leading logarithmic approximation, they depend on the choice of the
renormalization scheme (for a review, see \cite{Buras96}). The mesonic matrix
elements are customarily evaluated under the factorization hypothesis.
In order to ensure the $\mu$ and renormalization scheme independence for 
the physical amplitude, the matrix elements have to be computed in the same
renormalization scheme and renormalized at the same scale as $c_i(\mu)$. In
the naive factorization approach, the relevant Wilson coefficients for
external $W$-emission (or so-called ``class-I") and internal $W$-emission
(or ``class-II") nonpenguin amplitudes are given by $a_1=c_1+c_2/N_c$
and $a_2=c_2+c_1/N_c$, respectively, with $N_c$ the number of colors. 
However, it is well known that the factorization approximation does not 
work for color-suppressed class-II decay modes (for a review, see 
\cite{Cheng87}). This means that it is mandatory to take into account the
nonfactorizable effects, especially for class-II modes.
For $B\to PP$ or $VP$ decays ($P$: pseudoscalar meson, $V$: vector meson),
nonfactorizable contributions can be lumped into the effective parameters
$a_1$ and $a_2$ \cite{Cheng94,Cheng97,Kamal}:
\be
a_1^{\rm eff}=c_1(\mu)+c_2(\mu)\left({1\over N_c}+\chi_1(\mu)\right), \qquad
a_2^{\rm eff}=c_2(\mu)+c_1(\mu)\left({1\over N_c}+\chi_2(\mu)\right), 
\en
where $\chi_i$ are nonfactorizable terms and receive main contributions from
the color-octet current operators. Without loss of generality, the effective 
parameters $a_i^{\rm eff}$ include all the contributions
to the matrix elements and the nonfactorizable terms $\chi_i(\mu)$ take 
care of the correct $\mu$ dependence of the matrix elements \cite{Neubert}.
Schematically, we have
\be
\la c(\mu)O(\mu)\ra \to a^{\rm eff}\la O^{\rm tree}\ra_{\rm fact},
\en
where $\la O^{\rm tree}\ra_{\rm fact}$ is the matrix element of the tree
operator $O^{\rm tree}$ evaluated using the factorization method. Therefore,
$a_i^{\rm eff}$ are renormalization-scheme and -scale independent. Although
nonfactorizable contributions are nonperturbative in nature and thus
difficult to estimate, the effective
coefficients $a_i^{\rm eff}$ can be determined from experiment.
From $B\to D^{(*)}\pi(\rho)$ and $B\to J/\psi K^
{(*)}$ decays, we found that $a_1\sim (0.96-1.10)$ \cite{CT} and 
$a_2=0.26\sim 0.30$ \cite{Cheng97}. In the present analysis we assign the 
values
\be
a_1^{\rm eff}=1.02, \qquad a_2^{\rm eff}=0.28,
\en
for exclusive two-body decays of $B$ mesons.
By comparing (5) with (7), we learn that factorization works reasonably 
well for class-I decay modes, but nonfactorizable effects play an essential 
role for class-II channels.

  Unlike the effective coefficients $a_{1,2}^{\rm eff}$, which can be 
empirically
determined from experiment, {\it a priori} it is not clear how reliable is
the factorization approach for the matrix elements of the QCD and electroweak
penguin operators. Recently CLEO II has reported preliminary results of
charmless hadronic $B$ decays, namely, $B\to K^\pm\pi^\mp,~B\to K^0\pi
^\pm$ \cite{CLEO1,Alexander}, dominated by the penguin mechanism.
It turns out that the predictions based on factorization are in fair agreement
with data \cite{Fleischer97}. Although at present we are not able to determine
the analogous effective penguin coefficients that include nonfactorizable 
effects from experiment, we can nevertheless construct penguin Wilson 
coefficients that are
independent of the choice of the renormalization scale and the renormalization
scheme (for a review, see \cite{Fleischer97b}).
The one-loop QCD and electroweak corrections to matrix elements are 
parametrized by the matrices $\m_s$ and $\m_e$, respectively 
\cite{Buras92,Fleischer93}
\be
\la O_i(\mu)\ra=\left[\,{\rm I}+{\alpha_s(\mu)\over 4\pi}\m_s(\mu)+{\alpha
\over 4\pi}\m_e(\mu)\right]_{ij}\la O_j^{\rm tree}\ra.
\en
As a consequence,
\be
c_i(\mu)\la O_i(\mu)\ra=\,c_i(\mu)\left[\,{\rm I}+{\alpha_s(\mu)\over 4\pi}
\m_s(\mu)+{\alpha\over 4\pi}\m_e(\mu)\right]_{ij}\la O_j^{\rm tree}\ra\equiv
\tilde{c}_j\la O_j^{\rm tree}\ra.
\en
One-loop penguin matrix elements of the operators $O_{1,2}$ have been 
calculated with the results \cite{Buras92,Fleischer93} (penguin diagrams for
the matrix elements of penguin operators are calculated in \cite{Kramer}):
\be
&& \m_s(\mu)_{13}=\m_s(\mu)_{15}=-3\m_s(\mu)_{14}=-3\m_s(\mu)_{16}=\,{1\over 
6}\left[{2\over 3}\kappa+G(m_c,k,\mu)\right],   \non \\
&& \m_e(\mu)_{17}=\m_e(\mu)_{19}=3\m_e(\mu)_{27}=3\m_e(\mu)_{29}=\,-{4\over 9}
\left[{2\over 3}\kappa+G(m_c,k,\mu)\right],   
\en
where $\kappa$ parametrizes the renormalization scheme dependence. For 
example, for the naive dimension regularization (NDR) scheme and the 't 
Hooft-Veltman (HV) scheme, $\kappa$ is given by \cite{Fleischer93}
\be
\kappa=\cases{ -1 & (NDR),  \cr 0 & (HV).  \cr}
\en
The function $G(m_c,k,\mu)$ in Eq.~(10) defined by
\be
G(m_c,k,\mu)=-4\int_0^1 dx\,x(1-x)\ln\left[{m_c^2-k^2x(1-x)\over \mu^2}\right],
\en
has the analytic expression:
\be
&& {\rm Re}\,G=\,{2\over 3}\left(-\ln{m_c^2\over\mu^2}+{5\over 3}+4{m_c^2\over
k^2}-(1+2{m_c^2\over k^2})\sqrt{1-4{m_c^2\over k^2}}\,\ln\,{1+\sqrt{1-4{m_c^2
\over k^2}}\over 1-\sqrt{1-4{m_c^2\over k^2}}}\right),   \non \\
&& {\rm Im}\,G=\,{2\over 3}\pi\left(1+2{m_c^2\over k^2}\right)\sqrt{1-4{m_c^2
\over k^2}},
\en 
for $k^2>4m_c^2$,
where $k^2$ is the momentum squared of the virtual gluon. Note that the
imaginary part of $G(m,k,\mu)$, which furnishes the desired dynamic strong
phase necessary for generating CP asymmetries in charged $B$ decays,
arises only in the time-like penguin diagram
with $k^2>4m^2$. Eqs.~(9) and (10) lead to
the renormalization scale and scheme independent penguin Wilson coefficients:
\be
&& \tilde{c}_3=c_3(\mu)+{1\over 3}\tilde P_s(\mu), \qquad \tilde{c}_4=
c_4(\mu)-\tilde P_s(\mu), \qquad \tilde{c}_5=c_5(\mu)+{1\over 3}\tilde 
P_s(\mu), \non \\
&& \tilde{c}_6=c_6(\mu)-\tilde P_s(\mu),  \qquad \tilde{c}_{7(9)}=c_{7(9)}
(\mu)-\tilde P_e(\mu), \qquad \tilde{c}_{8(10)}=c_{8(10)},
\en
where
\be
\tilde P_s(\mu) &=& {\alpha_s(\mu)\over 8\pi}\left[{2\over 3}\kappa+G(m_c,k,
\mu)\right]c_1(\mu),   \non \\
\tilde P_e(\mu) &=& {\alpha\over 9\pi}\left[{2\over 3}\kappa+G(m_c,k,\mu)
\right][c_1(\mu)+3c_2(\mu)].
\en

   As noted before, the Wilson coefficients $c_i(\mu)$ have been evaluated 
to the next-to-leading order in the NDR, HV and RI (regularization independent)
renormalization schemes \cite{Buras96,Ciuchini}. For the purpose of 
illustration, we use the $\Delta B=1$ Wilson coefficients obtained in 
HV and NDR schemes at $\mu=4.4$ GeV, $\Lambda^{(5)}_{\overline{\rm MS}}=225$ 
MeV and $m_t=170$ GeV in Table 22 of \cite{Buras96}. From Eqs.~(11-15)
we find that $\tilde{c}_3-\tilde{c}_{10}$ obtained from $c_i(\mu)$ in the 
HV and NDR 
schemes are numerically very similar, implying the renormalization scheme
independence of $\tilde c_i$. The values of $\tilde c_i$ at $k^2=
m^2_b/2$ are
\be
&& \tilde{c}_3=0.0200+i0.0048, \qquad \qquad \quad\tilde{c}_4=-0.0515-i0.0143, 
\non  \\
&& \tilde{c}_5=0.0155+i0.0048, \qquad \qquad \quad\tilde{c}_6=-0.0565-i0.0143, 
\non  \\
&& \tilde{c}_7=-(0.0757+i0.0558)\alpha, \qquad \quad \tilde{c}_8=0.057\,
\alpha, \non  \\
&& \tilde{c}_9=-(1.3648+i0.0558)\alpha, \qquad \quad \tilde{c}_{10}=
0.264\,\alpha.
\en
One can also obtain $\tilde c_i$ by first evaluating the renormalization 
scheme independent but $\mu$ dependent Wilson coefficients $\bar c_i(\mu)$,
and then connecting them to $\tilde c_i$ via the relation
\be
\tilde c_i=\left[\,{\rm I}+{\alpha_s(\mu)\over 4\pi}(\m_s(\mu)-\hat r_s(\mu))
+{\alpha \over 4\pi}(\m_e(\mu)-\hat r_e(\mu))\right]_{ji}\bar c_j(\mu),
\en
where the expressions of the matrices $\hat r_{s,e}(\mu)$ are given in
\cite{Buras92,Fleischer93}. Using the values of $\bar c_i(\mu)$ obtained 
at $\mu=m_b=5$ GeV in \cite{Desh},
the reader can check that $\tilde c_3-\tilde{c}_{10}$ obtained in this way are 
numerically in accordance with (16) except for a small discrepancy with
$\tilde c_7$. This shows the renormalization scale $\mu$ independence of 
the Wilson coefficients $\tilde c_i$.

   Two remarks are in order. (i) One can equally use the Wilson coefficients 
$c_i(\mu)$ or the renormalization-scheme independent $\bar c_i(\mu)$ to
calculate the physical amplitude. However, the matrix elements of four-quark
operators must be renormalized at the same scale $\mu$ and evaluated in
the same renormalization scheme for the latter. The great advantage of working
with the renormalization-scale and -scheme independent $a_{1,2}^{\rm eff}$ 
and $\tilde{c}_3-\tilde{c}_{10}$ is two-fold: First, the corresponding 
tree-level
operators are scale and scheme independent and hence they can be evaluated 
using factorization. Second, long-distance nonfactorizable effects for 
external
and internal $W$-emissions are already included in the effective coefficients 
$a_{1,2}^{\rm eff}$. 
(ii) It is instructive to compare $\tilde{c}_i$ with the leading-order
Wilson coefficients $c_i(\mu)$ evaluated at $\mu=4.4$ GeV and $\Lambda_{
\overline{\rm MS}}^{(5)}=225$ MeV (see Table 22 of \cite{Buras96}):
\be
&& c_3=0.014, \quad c_4=-0.030, \quad c_5=0.009, \quad c_6=-0.038,   \non \\
&& c_7=0.045\,\alpha, \quad c_8=0.048\,\alpha, \quad c_9=-1.280\,\alpha, \quad 
c_{10}=0.328\,\alpha.
\en
It is evident that next-to-leading order corrections to the Wilson 
coefficients of QCD penguin operators are quite sizable.

\medskip
{\bf 3.} We are ready to employ the weak Hamiltonian (1) and the factorization
approximation to calculate exclusive charmless rare $B$ decays. We take the 
decay $B^-\to\eta' K^-$ as an example. Factorization leads to
\be
A(B^-\to\eta' K^-) &=& {G_F\over\sqrt{2}}\Bigg\{ V_{ub}V_{us}^*\left(a_1^{\rm 
eff}X_1+a_2^{\rm eff}X_{2u}+a_1^{\rm eff}X_3\right)+V_{cb}V_{cs}^*a_2^{\rm 
eff}X_{2u}(f_{\eta'}^{(\bar cc)}/f_{\eta'}^{(\bar uu)})    \non \\
&-& V_{tb}V_{ts}^*\Bigg[ \left(a_4+a_{10}+2(a_6+a_8){m_K^2\over (m_s+m_u)(m_b
-m_u)}\right)X_1   \non \\
&+& \left(2a_3-2a_5-{1\over 2}a_7+{1\over 2}a_9\right)X_{2u}   \non \\
&+& \left(a_3+a_4-a_5+{1\over 2}a_7-{1\over 2}a_9-{1\over 2}a_{10}+(a_6-{
1\over 2}a_8){m^2_{\eta'}\over m_s(m_b-m_s)}\right)X_{2s}   \non \\
&+& \left(a_4+a_{10}+2(a_6+a_8){m_B^2\over (m_s-m_u)(m_b+m_u)}\right)X_3
\Bigg]\Bigg\},
\en
where $a_{2i}\equiv\tilde{c}_{2i}+{1\over N_c}\tilde{c}_{2i-1},~a_{2i-1}\equiv
\tilde{c}_{2i-1}+{1\over N_c}\tilde{c}_{2i}$ ($N_c=3$) for $i\geq 2$, 
$X_i$ are factorizable terms:
\be
X_1 &\equiv& \la K^-|(\bar su)_\vma|0\ra\la\eta'|(\bar ub)_\vma|B^-\ra=i
f_K(m_B^2-m^2_{\eta'})F_0^{B\eta'}(m_K^2),   \non \\
X_{2q} &\equiv& \la \eta'|(\bar qq)_\vma|0\ra\la K^-|(\bar sb)_\vma|B^-\ra=
if_{\eta'}^{(\bar qq)}(m_B^2-m^2_K)F_0^{BK}(m_{\eta'}^2),   \non \\
X_3 &\equiv& \la\eta'K^-|(\bar su)_\vma|0\ra\la 0|(\bar ub)_\vma|B^-\ra,
\en
and $F_0$ is the form factor defined in \cite{BSW}. In the derivation of
Eq.~(19) we have applied the isospin relation $X_{2d}=X_{2u}$ and
equations of motion for $(S-P)(S+P)$ penguin matrix elements. 
The wave functions of the physical $\eta'$ and $\eta$ states are related to
that of the SU(3) singlet state $\eta_0$ and octet state $\eta_8$ by
\be
\eta'=\eta_8\sin\theta+\eta_0\cos\theta, \qquad \eta=\eta_8\cos\theta-\eta_0
\sin\theta
\en
with $\theta\approx -20^\circ$. When the $\eta-\eta'$ mixing angle is 
$-19.5^\circ$, 
the $\eta'$ and $\eta$ wave functions have simple expressions \cite{Chau1}
\be
|\eta'\ra={1\over\sqrt{6}}|\bar uu+\bar dd+2\bar ss\ra, \qquad
|\eta\ra={1\over\sqrt{3}}|\bar uu+\bar dd-\bar ss\ra.
\en
For simplicity we will employ $\theta=-19.5^\circ$ in ensuing discussion.
In SU(3) limit, the $\eta'$ and $\eta$ decay constants become \cite{Chau1}
\be
f_{\eta'}^{(\bar uu)}=f_{\eta'}^{(\bar dd)}=f_{\eta'}^{(\bar ss)}/2=
f_\pi/\sqrt{6}=54\,{\rm MeV},   \non \\
f_{\eta}^{(\bar uu)}=f_{\eta}^{(\bar dd)}=-f_{\eta}^{(\bar ss)}=
f_\pi/\sqrt{3}=77\,{\rm MeV}.
\en
The measured physical decay constants $f_\eta$ and $f_{\eta'}$ are fairly 
close to $f_\pi$ \cite{PDG}.
As a result, in SU(3) limit, $X_{2s}=2X_{2u}$ for $\eta'$ production and 
$X_{2s}=-X_{2u}$ for the $\eta$. In the limit of SU(3)-flavor symmetry, we
also have
\be
F_0^{BK}(0)=\sqrt{6}F_0^{B\eta'}(0)=F_0^{B\pi^\pm}(0),
\en
where the factor of $\sqrt{6}$ comes from the normalization constant of the 
$\eta'$ wave function [see Eq.~(22)].

   For non-penguin external and internal $W$-emission contributions to
$B^\pm\to \eta' K^\pm$, we employ the effective coefficients $a_{1,2}
^{\rm eff}$ [Eq.~(7)] since they take into account nonfactorizable effects.
In Eq.~(19) $X_3$ stands for the $W$-annihilation contribution. In the
penguin mechanism, $X_3$ arises from the space-like penguin diagram. It is 
common to argue that $W$-annihilation is negligible due to helicity 
suppression, corresponding to form factor suppression at large momentum 
transfer, $q^2=m_B^2$ (for a recent study, see \cite{Xing}).
However, we see from Eq.~(19) that it is largely enhanced in the space-like 
penguin diagram by a factor of $m_B^2/(m_bm_s)$. Unfortunately, we do not 
have a reliable method for estimating $W$-annihilation, though recent 
PQCD calculations suggest that space-like penguins are small compared to 
the time-like ones \cite{Lu}.

   In terms of the Wolfenstein parametrization \cite{Wol}, the relevant 
quark mixing matrix elements for $B\to\eta' K$ are
\be
V_{ub}V_{us}^*=A\lambda^4(\rho-i\eta), \quad V_{cb}V_{cs}^*=A\lambda^2(1-
{\lambda^2\over 2}), \quad V_{tb}V_{ts}^*=-A\lambda^2,
\en
where $\lambda=0.22$ and $A=0.804$ for $|V_{cb}|=0.039$. Since the parameters 
$\rho$ and $\eta$ have not been measured separately, it is desirable to have 
the results presented for allowed regions of $\rho$ and $\eta$. As 
$\rho$ and $\eta$ are approximately constrained by $0.2<\eta<0.4$ and
$-0.3<\rho<0.3$ (see, e.g., \cite{Buras96}), we will fix $\eta\sim 0.30$ and
take $\rho=0.30$ as well as $\rho=-0.30$. We also need input of other 
parameters.
For quark masses we use $m_u=5$ MeV, $m_d=10$ MeV, $m_s=175$ MeV, $m_c=1.5$
GeV and $m_b=5.0$ GeV. Heavy-to-light mesonic form factors have been 
evaluated in the relativistic quark model \cite{BSW,CCW}. To be specific, 
we use $F_0^{B\eta'}(0)=0.254/\sqrt{6}$ \cite{BSW}
\footnote{It should be stressed that the form factors given in \cite{BSW} 
for $B(D)\to\pi^0,~\eta,~\eta'$ transitions do not take into account the 
normalization constant of the neutral meson wave functions.}
and $F_0^{BK}(0)=0.34$ \cite{CCW}. For the lifetime of $B$ mesons, the
world average is $\tau(B_d)=(1.55\pm 0.04)$ ps and $\tau(B^\pm)=(1.66\pm 
0.04)$ ps \cite{Richman}. Since what CLEO has measured is the combined 
branching ratio 
\be
{\cal B}(B^\pm\to\eta' K^\pm)\equiv {1\over 2}\left[{\cal B}(B^+\to\eta' K^+)
+{\cal B}(B^-\to\eta' K^-)\right],
\en
we shall present the results for the combined one. Neglecting the 
$W$-annihilation
contribution characterized by the parameter $X_3$, we find that in the absence
of the transition $c\bar c\to\eta'$ (see Table I)
\footnote{The prediction ${\cal B}(B^\pm\to\eta' K^\pm)=3.6\times 10^{-5}$
given in \cite{Chau1}
is too large by about a factor of 2 because the normalization constant (i.e.
$1/\sqrt{6}$) of the $\eta'$ wave function was not taken into account in the 
form factor $F_0^{B\eta'}$ (see also the previous footnote).}
\be
{\cal B}(B^\pm\to\eta' K^\pm)=(1.4-1.8)\times 10^{-5}\quad {\rm for}~\fp=0,
\en
which is substantially smaller than the CLEO measurement (2). 
It is worth
emphasizing that if the Wilson coefficients are evaluated only to the leading
order [cf. Eq.~(18)], then the resultant predictions become even worse:
${\cal B}(B^\pm\to\eta' K^\pm)_{\rm L.O.}=(0.5-0.8)\times 10^{-5}$. 
In fact, the leading order
calculation is already ruled out by the recent CLEO measurement of $B^\pm
\to\pi^\pm K$ and $B_d\to\pi^\mp K^\pm$ \cite{Fleischer97}.

   We briefly discuss the uncertainties associated with the prediction of
${\cal B}(B^\pm\to\eta' K^\pm)$ and the numerical results given in Table I. 
We have neglected $W$-annihilation, space-like penguin diagrams and
nonfactorizable contributions to the matrix elements of penguin operators; 
all of them are difficult to estimate. Other major sources of uncertainties 
come from the choice of form factors and light quark masses. 
It is natural to ask if the CLEO result for $B^\pm\to\eta' K^\pm$ can be 
explained by different choices of the strange quark mass $m_s$ and form
factors. For a given set of form factors, it is easily seen that the branching
ratio increases with decreasing $m_s$. For example, we find a large
branching ratio ${\cal B}(B^\pm\to\eta' K^\pm)=6.5\times 10^{-5}$ for
$m_s=60$ MeV and $\rho<0$.
However, the use of a small strange quark mass is in conflict with
a recent measurement in $\tau$ decay, $m_s(m_\tau)=212^{+30}_{-35}$ MeV
\cite{Davier}. It is also true that the measured branching ratio of the
$\eta' K$ mode can be accommodated by choosing large form factors, e.g.,
$F_0^{BK}(0)\approx \sqrt{6}F_0^{B\eta'}(0)\approx 0.61$. However, this will 
break the SU(3)-symmetry relation (24) very badly. All the existing 
QCD sum rule, lattice and quark model calculations indicate that $F_0^{B
\pi^\pm}(0)\approx 0.33$ or less (for a review, see \cite{Cas}). The recent
observed charmless decays $B\to K^\pm\pi^\mp,~B\to K^0\pi^\pm$ dominated by
the penguin diagram are governed by the form factor $F_0^{B\pi^\pm}$. We
find that the predictions for $B\to K\pi$ based on the standard approach
with $F_0^{B\pi^\pm}(0)=0.33$
are in agreement with data. Therefore, in view of the SU(3) relation (24)
for form factors, it is very unlikely that $F_0^{BK}(0)$ and $\sqrt{6}F_0^{B
\eta'}(0)$ can deviate much from 0.33.
We should also stress that it is dangerous to adjust the form factors in order
to fit a few particular modes; the comparison
between theory and experiment should be done using the same set of form 
factors for all channels. We thus conclude that the large branching ratio
of $B^\pm\to\eta' K^\pm$ observed by CLEO cannot be explained by the
conventional approach.

   The sizable discrepancy between theory (27) and experiment (2) strongly
suggests a new mechanism unique to the $\eta'$ meson. It appears that the 
most natural one is the Cabibbo-allowed process $b\to c\bar c s$ followed 
by a conversion of the $c\bar c$ pair into the $\eta'$ via two gluon 
exchanges [see Eq.~(19)]. This new internal $W$-emission contribution
is important since its mixing angle $V_{cb}V_{cs}^*$ is as large as that
of the penguin amplitude and yet its Wilson coefficient $a_2^{\rm eff}$ 
is larger than that of penguin operators. 
The decay constant $f_{\eta'}^{(\bar cc)}$, 
defined as $\la 0|\bar c\gamma_\mu\gamma_5c|\eta'\ra=if_{\eta'}^{(\bar cc)}
q_\mu$, can be reduced via the OPE to a matrix element of a particular 
dimension-6 pseudoscalar gluonic operator \cite{Halp2}. It has been estimated 
in \cite{Halp2}, based on the OPE, large-$N_c$ approach and QCD low energy 
theorems, that $|f_{\eta'}^{(\bar cc)}|=(50-180)$ MeV. (An independent
estimate of $\fp$ based on the instanton-liquid model is given in 
\cite{Shur}.) It was claimed in \cite{Halp2,Halp1} 
that $|f_{\eta'}^{(\bar cc)}|\sim 140$ MeV is needed in order to exhaust the 
CLEO observation of $B^\pm\to \eta' K^\pm$ and $B\to\eta'+X$ by the mechanism
$b\to c\bar c+s\to\eta'+s$ via gluon exchanges.
In view of the fact that $f_{\eta'}^{(\bar uu)}$ is only of order 50 MeV,
a large value of $\fp$ seems to be very unlikely. We believe that it is more
plausible to take the lower side of the theoretical estimate, namely
$|f_{\eta'}^{(\bar cc)}|\sim 50\,{\rm MeV}$.
\footnote{The mixing angle of the $\eta'$ with the $\bar cc$ state is of 
order $7^\circ$ for $|f_{\eta'}^{(\bar cc)}|\sim 50\,{\rm MeV}$.}
On the other hand, it is claimed in \cite{Chao} that $|f_{\eta'}^{(\bar cc)}|
\leq 40$ MeV. From Eq.~(19) it is
clear that this additional contribution will enhance the decay rate of
$B^\pm\to\eta' K^\pm$ provided that the sign of $\fp$ is opposite to that of 
$f_{\eta'}^{(\bar uu)}$. (Note that the estimate of $\fp$ in \cite{Halp2}
does not fix its sign.) For $f_{\eta'}^{(\bar cc)}=-50$ MeV, we obtain
\be
{\cal B}(B^\pm\to\eta' K^\pm) &=& (5.8-6.7)\times 10^{-5},  \non \\  
{\cal B}(\stackrel{_{_{(-)}}}{B}\!\!{^0}\to\eta' \stackrel{_{_{(-)}}}{K}\!\!
{^0}) &=& (5.7-5.9)\times 10^{-5},
\en
which are in agreement with the CLEO measurement (2), especially
when $\rho$ is negative. Contrary to what has been claimed in \cite{Halp2}, 
it is not necessary to introduce a 
large value of $\fp$ to account for the experimental observation of
$B^\pm\to\eta' K^\pm$.

    Suppose the dominant mechanism for the observed large inclusive $B^\pm
\to\eta'+X$ signal comes either from the $b\to s\,g^*$ penguin followed by 
the transition $g^*\to g\,\eta'$ via the QCD anomaly, or from the process 
$b\to 
(c\bar c)_8+s$, followed by the conversion $(c\bar c)_8\to\eta'+X$. Will
these mechanisms still play an important role in exclusive $B^\pm\to\eta'
K^\pm$ decays ? It is easily seen that since the above two mechanisms
involve a production of a gluon before hadronization, they will
not make contributions to two-body exclusive decays unless the gluon is soft
and absorbed in the wave function of the $\eta'$. Another possibility 
is the penguin-like process $b\to s+g^*g^*\to s+\eta'$, but it is a higher
order penguin mechanism.

   The factorizable contribution to $B^-\to\eta K^-$ has a similar
expression as (19) for $B^-\to\eta' K^-$ with the replacement
\be
\fp \to f_\eta^{(\bar cc)}\approx -\sin\theta\fp,
\en
where $\theta\approx -20^\circ$ is the $\eta-\eta'$ mixing angle. Because
of the destructive interference in the penguin diagrams due to the relation
$X_{2s}=-X_{2u}$ in SU(3) limit and because of the smallness of the $c\bar c
\to\eta_0$ contribution to $\eta$ production, the decay $B\to
\eta K$ is suppressed. We see from Table I that the branching ratio
of $B\to\eta K$ is smaller than that of $B\to\eta' K$
by an order of magnitude. Nevertheless, a measurement of ${\cal B}(
B\to\eta K)$ of order $(2-5)\times 10^{-6}$ will confirm the 
importance of the $c\bar c\to\eta'$ mechanism. From Table I we see that 
the electroweak penguin effects are in general very small, but they
become important for $B\to\eta K$ and $B\to\eta K^*$ decays due to a large
cancellation of QCD penguin contributions in these decay modes. It is
worth stressing that CP asymmetries in $B^\pm\to\eta(\eta')K^\pm(K^{*\pm})$
decays will be largely diluted in the presence of the $\bar cc$ conversion
into the $\eta'$ \cite{CB97}.

   The mechanism of $c\bar c$ conversion into $\eta'$, if correct, is 
probably most dramatic in the decay $B\to\eta' K^{*}$. 
The factorization amplitude for $B^-\to\eta' K^{*-}$ is given by
\be
A(B^-\to\eta' K^{*-}) &=& {G_F\over\sqrt{2}}\Bigg\{ V_{ub}V_{us}^*\left(a_1^
{\rm eff}X'_1+a_2^{\rm eff}X'_{2u}+a_1^{\rm eff}X'_3\right)+V_{cb}V_{cs}^*
a_2^{\rm eff}X'_{2u}(f_{\eta'}^{(\bar cc)}/f_{\eta'}^{(\bar uu)})    \non \\
&-& V_{tb}V_{ts}^*\Bigg[(a_4+a_{10})X'_1 +\left(2a_3-2a_5-{1\over 2}a_7+
{1\over 2}a_9\right)X'_{2u}   \non \\
&+& \left(a_3+a_4-a_5+{1\over 2}a_7-{1\over 2}a_9-{1\over 2}a_{10}-(a_6-{
1\over 2}a_8){m^2_{\eta'}\over m_s(m_b+m_s)}\right)X'_{2s}   \non \\
&+& \left(a_4+a_{10}-2(a_6+a_8){m_B^2\over (m_s+m_u)(m_b+m_u)}\right)X'_3
\Bigg]\Bigg\},
\en
where
\be
X'_1 &\equiv& \la K^{*-}|(\bar su)_\vma|0\ra\la\eta'|(\bar ub)_\vma|B^-\ra=-
2f_{K^*}m_{K^*}F_1^{B\eta'}(m_{K^*}^2)(\ep\cdot p_B),   
\non \\
X'_{2q} &\equiv& \la\eta'|(\bar qq)_\vma|0\ra\la K^{*-}|(\bar sb)_\vma|B^-\ra=
-2f_{\eta'}^{(\bar qq)}m_{K^*}A_0^{BK^*}(m_{\eta'}^2)(\ep\cdot p_B), 
\non \\
X'_3 &\equiv& \la\eta'K^{*-}|(\bar su)_\vma|0\ra\la 0|(\bar ub)_\vma|B^-\ra.
\en
Note that, contrary to $B\to\eta' K$ decay,
there is no contribution proportional to $X'_1$ arising from the 
$(S-P)(S+P)$ part of the penguin operators $O_6$ and $O_8$. The $q^2$ 
dependence
of the form factors $F_0,~F_1,~A_0$, defined in \cite{BSW}, can be calculated
in the framework of the relativistic quark model. A direct calculation of
$B\to P$ and $B\to V$ form factors at time-like momentum transfer in the
relativistic light-front quark model just became available recently 
\cite{CCW}. It is found that $A_0,~F_1$ exhibit a dipole behavior, while
$F_0$ shows a monopole dependence. At $q^2=0$, the relevant form factors
are \cite{CCW}
\be
F_1^{BK}(0)=0.34, \quad F_1^{B\pi}(0)=0.26, \quad A_0^{BK^*}(0)=0.32, \quad
A_0^{B\rho}(0)=0.28.
\en
We then utilize the above-mentioned $q^2$ behavior to compute the form factors
at the desired momentum transfer squared. Owing to the small penguin
contributions, the standard approach's estimate of ${\cal B}(B\to\eta' 
K^{*})$ is small, of order $(2-5)\times 10^{-7}$ for $B^\pm\to\eta' K^{*\pm}$
and $1\times 10^{-7}$ for $B^0\to\eta' K^{*0}$.
Since the branching ratio
due to the process $b\to c\bar c+s\to\eta'+s$ is of order $10^{-5}$, 
${\cal B}(B\to\eta' K^{*})$ is greatly enhanced to 
the order of $(1-2)\times 10^{-5}$. It has been argued in
\cite{Halp1} that ${\cal B}(B\to\eta' K^*)$ is about twice larger than that
of $B\to\eta' K$, which is certainly not the case in our calculation.
It is interesting to note that the ratio ${\cal B}(B\to\eta' K^{*})/
{\cal B}(B\to\eta K^{*})$ is much less than unity in the naive
estimate \cite{Lipkin}, but it becomes greater than unity in the presence 
of $c\bar c$ conversion into the $\eta_0$ (see Table I).

\vskip 0.4cm
\begin{table}
{{\small Table I. Averaged branching ratios for charmless $B$ decays to 
$\eta'$ and $\eta$ [see Eq.~(26)], where ``Tree" refers to branching ratios 
from 
tree diagrams only, ``Tree+QCD" from tree and QCD penguin diagrams, and 
``Full" denotes full contributions from tree, QCD and electroweak (EW) penguin
diagrams in conjunction with contributions from the process $c\bar 
c\to\eta_0$. Predictions are for $k^2=m_b^2/2$, $\fp=-50$ MeV, 
$\eta=0.30,~\rho=0.30$ (the first number in parentheses) and $\rho=-0.30$ 
(the second number in parentheses).}
{\footnotesize
\begin{center}
\begin{tabular}{|l|c c c c |c|} \hline
Decay & Tree & Tree$+$QCD & Tree$+$QCD$+$EW & Full & Exp. \cite{CLEO1,CLEO2}\\ 
\hline 
$B^\pm\to\eta' K^\pm$ & $1.45\times 10^{-7}$ & $(1.40,~1.82)\,10^{-5}$ & 
$(1.35,~1.76)\,10^{-5}$ & $(5.83,~6.67)\,10^{-5}$ &
$(7.1^{+2.5}_{-2.1}\pm 0.9)\,10^{-5}$ \\
$B^\pm\to\eta K^\pm$ & $4.12\times 10^{-7}$ & $(2.73,~6.54)\,10^{-7}$ &
$(0.21,~1.05)\,10^{-6}$ & $(1.94,\,5.15)\,10^{-6}$  & $<0.8\times 10^{-5}$ \\
$B^\pm\to\eta' K^{*\pm}$ & $2.10\times 10^{-7}$ & $(1.57,~5.23)\,10^{-7}$ &
$(1.67,~4.57)\,10^{-7}$ & $(1.14,~1.62)\,10^{-5}$ & $<29\times 10^{-5}$ \\
$B^\pm\to\eta K^{*\pm}$ & $6.23\times 10^{-7}$ & $(0.41,~2.09)\,10^{-6}$ &
$(0.56,~2.79)\,10^{-6}$ & $(2.95,\,7.79)\,10^{-6}$ & $<24\times 10^{-5}$ \\
$B^\pm\to\eta'\pi^{\pm}$ & $1.88\times 10^{-6}$ & $(8.94,~8.13)\,10^{-6}$ &
$(8.91,~8.09)\,10^{-6}$ & $(1.36,~1.24)\, 10^{-5}$ & $<4.5\times 10^{-5}$ \\
$B^\pm\to\eta \pi^{\pm}$ & $5.34\times 10^{-6}$ & $(9.19,~1.89)\,10^{-6}$ &
$(9.32,~1.91)\,10^{-6}$ & $(1.07,\,0.22)\, 10^{-5}$ & $<0.8\times 10^{-5}$ \\
$B^\pm\to\eta' \rho^{\pm}$ & $3.90\times 10^{-6}$ & $(0.45,~2.37)\,10^{-5}$ &
$(0.45,~2.39)\,10^{-5}$ & $(0.47,~1.83)\, 10^{-5}$ & \\
$B^\pm\to\eta \rho^{\pm}$ & $1.16\times 10^{-5}$ & $(1.05,~1.84)\,10^{-5}$ &
$(1.06,~1.80)\,10^{-5}$ & $(1.15,\,1.65)\, 10^{-5}$ &  \\
\hline
$ B_d\to\eta' K^0$ & $7.93\times 10^{-9}$ &$(1.45,\,1.54)\, 10^{-5}$
&$(1.40,\,1.49)\, 10^{-5}$ &$(5.73,\,5.92)\, 10^{-5}$ & $(5.3^{+2.8}_{-2.2}
\pm 1.2)\, 10^{-5}$ \\
$ B_d\to\eta K^0$  & $1.58\times 10^{-8}$ & $(0.31,\,1.07)\, 10^{-7}$ 
& $(1.52,\,3.14)\, 10^{-7}$ & $(2.67,\,3.29)\, 10^{-6}$ & \\
$ B_d\to\eta' K^{*0}$ & $6.05\times 10^{-9}$ & $(0.93,\,1.50)\,
10^{-7}$ & $(0.73,\,1.17)\, 10^{-7}$ & $(1.28,\,1.36)\, 10^{-5}$ &
$<9.9\times10^{-5}$ \\ 
$ B_d\to\eta K^{*0}$ & $1.32\times 10^{-8}$ & $(4.56,\,6.88)\,
10^{-7}$  & $(0.82,\,1.13)\, 10^{-6}$  & $(4.17,\,4.86)\, 10^{-6}$  &
$<3.3\times 10^{-5}$ \\ 
$ B_d\to\eta'\pi^{0}$ & $0.34\times 10^{-9}$ & $(2.27,\,6.81)\,
10^{-6}$  & $(2.20,\,6.61)\, 10^{-6}$  & $(3.72,\,9.19)\, 10^{-6}$  & 
$<2.2\times 10^{-5}$ \\
$ B_d\to\eta \pi^{0}$ & $1.98\times 10^{-9}$ & $(0.73,\,2.43)\,
10^{-6}$ & $(0.71,\,2.36)\, 10^{-6}$ & $(0.96,\,2.83)\, 10^{-6}$ & \\
$ B_d\to\eta'\rho^{0}$ & $1.03\times 10^{-8}$ &$(1.57,\,4.18)\,
10^{-6}$ &$(1.67,\,4.45)\, 10^{-6}$ &$(0.77,\,2.78)\, 10^{-6}$ & \\
$ B_d\to\eta \rho^{0}$ & $5.86\times 10^{-8}$ &$(2.84,\,2.60)\,
10^{-7}$ &$(3.20,\,3.24)\, 10^{-7}$ &$(1.67,\,1.74)\, 10^{-7}$ &
$<8.4\times 10^{-5}$   \\
\hline
\end{tabular}
\end{center} } }
\end{table} 
\vskip 0.4cm

  For $B\to\eta'(\eta)\pi(\rho)$ decays, the mechanism of $c\bar c\to\eta_0$
is less dramatic since it does not gain mixing-angle enhancement as in 
the case of $B\to\eta'(\eta)K(K^*)$. We see from Table I that the full
branching ratios in general are close to the predictions in the conventional
approach, indicating the minor role played by the charm content of the 
$\eta'$. Three remarks are in order. First, there is a large enhancement
of order $m^2_{\eta'(\eta)}/(m_qm_b)~(q=u,d)$ occurred in the matrix 
elements of $(S-P)(S+P)$ penguin operators. The calculation in this case is 
thus sensitive to the light quark masses $m_u$ and $m_d$. Second, the 
branching ratio of $B^\pm\to\eta\pi^\pm$ is predicted to be $1\times 
10^{-5}$ for positive 
$\rho$, while experimentally ${\cal B}(B^\pm\to\eta h^\pm)<8.0\times 10^{-6}$
for $h=\pi,~K$ \cite{CLEO1}.
This again indicates that a negative $\rho$ is preferable.
Third, the decays $B^\pm\to\eta\rho^\pm$ are overwhelmly dominated by 
tree diagrams.

\medskip
{\bf 4.}~~To summarize, in view of the recent unexpectedly large branching 
ratios for
inclusive and exclusive charmless $B$ decays to $\eta'$, we have analyzed
the exclusive $B$ decays to $\eta'$ and $\eta$ in detail. We showed that the 
process $b\to c\bar c +s\to\eta'+s$ via gluon exchanges with 
$\fp\sim -50$ MeV is adequate to
explain the discrepancy between theory and experiment. A measurement of
${\cal B}(B\to\eta' K^*)$ of order $10^{-5}$ will give a strong support
of this gluon mechanism of $\eta'$ production. Moreover, we found that the 
decay rate of $B\to\eta' K^*$ is larger than $B\to\eta K^*$, contrary to
what expected from the standard approach. The predicted branching ratio
of $B^\pm\to\eta\pi^\pm$ for positive $\rho$ is marginally ruled out by 
experiment, implying that a negative $\rho$ is preferable.

\bigskip
\noindent ACKNOWLEDGMENT:~~This work was supported in part by the National 
Science Council of ROC under Contract Nos. NSC86-2112-M-001-020 and
NSC86-2112-M-001-010-Y.

\bigskip
\noindent {\it Note added}: The exclusive decays of $B$ to $\eta'$ and $\eta$ 
were also discussed recently by A. Ali and C. Greub (hep-ph/9707251), A. 
Datta, X.G. He and S. Pakvasa
(hep-ph/9707259) with conclusions different from ours.

\renewcommand{\baselinestretch}{1.1}
\newcommand{\bi}{\bibitem}
%

\newpage

\end{document}